\documentclass[twocolumn,aps,showpacs,pra,notitlepage]{revtex4-1}
\usepackage{url,psfrag,graphicx}
\usepackage{amsmath,amssymb}
\usepackage{bm}
\usepackage{hyperref}
\usepackage{epsfig,color}
\usepackage[section]{placeins}

\def\<{\left<}
\def\>{\right>}
\newcommand{\ket}[1]{|{#1}\rangle}

\def\elem<#1|#2|#3>{\left<#1\right|#2\left|#3\right>}
\def\({\left(}
\def\){\right)}

\begin{document}

\title{Quantum simulation of non-trivial topology}

\author{Octavi Boada}
\affiliation{Physics of Information Group, Instituto de Telecomunica\c{c}\~oes, P-1049-001 Lisbon, Portugal}

\author{Alessio Celi}
\affiliation{ICFO-Institut de Ci\`encies Fot\`oniques, Barcelona, Spain}

\author{Javier Rodr\'{\i}guez-Laguna} 
\affiliation{IFT-Instituto de F\'{\i}sica Te\'orica, UAM-CSIC, Madrid, Spain}
\affiliation{ICFO-Institut de Ci\`encies Fot\`oniques, Barcelona, Spain}
 
\author{Jos\'e I. Latorre}
\affiliation{Dept. ECM, Universitat de Barcelona, Spain}
 
\author{Maciej Lewenstein}
\affiliation{ICFO-Institut de Ci\`encies Fot\`oniques, Barcelona, Spain}

\bigskip

\begin{abstract}
We propose several designs to simulate quantum many-body systems in
manifolds with a non-trivial topology. The key idea is to create a
synthetic lattice combining real-space and internal degrees of freedom
via a suitable use of induced hoppings. The simplest example is the
conversion of an open spin-ladder into a closed spin-chain with
arbitrary boundary conditions. Further exploitation of the idea leads
to the conversion of open chains with internal degrees of freedom into
artificial tori and M\"obius strips of different kinds. We show that
in synthetic lattices the Hubbard model on sharp and scalable
manifolds with non-Euclidean topologies may be realized. We provide a
few examples of the effect that a change of topology can have on quantum systems 
amenable to simulation, both at the single-particle and at the many-body level.
\end{abstract}


\maketitle


\section{Introduction}

The research field of quantum simulation explores, among other goals,
the possibility of using well-controlled quantum systems to simulate
the behavior of other quantum systems whose dynamics escapes standard
theoretical or experimental approaches. As a relevant example, quantum
simulators have been used to successfully analyse condensed matter
phenomena \cite{lewenstein2012ultracold}. Through synthetic gauge
fields \cite{dalibard2011colloquium}, more ambitious and
multidisciplinary problems can be addressed, such as the determination of the
phase diagram of (lattice) gauge theories \cite{zohar2012simulating,
  banerjee2012atomic, tagliacozzo2013optical, zohar2013cold,
  banerjee2013atomic, tagliacozzo2013simulation}. This theoretical progress is
supported by vigorous experimental developments with a growing number
of platforms available for quantum simulation like cold neutral atoms
and molecules \cite{bloch2012quantum}, trapped ions
\cite{blatt2012quantum}, photonic crystals \cite{aspuru2012photonic},
NV-centers \cite{cai2013large}, and superconducting qubits
\cite{houck2012chip}.

On a different line of research, topological models have attracted
great interest as well. Topology is a key feature to understand many
physical phenomena, such as the quantum Hall and quantum spin-Hall
effects \cite{haldane1988model}, quantization of Dirac monopole charge
\cite{sakurai2011modern}, charge fractionalization and
non-perturbative properties of vacua of Yang-Mills theories
\cite{su1979solitons,jackiw1976solitons,witten1979current,veneziano1979u},
etc. Topology also plays an essential role in engineering novel states
of ultracold matter, such as topological insulators
\cite{hasan2010colloquium}. Notably, topological protection has been
considered as a resource for quantum computation \cite{nayak2008non}.
Nonetheless, non-trivial topology is not easy to implement in
practical systems. For instance, there is no obvious way to manipulate
a 2D condensed matter system to be topologically connected as on a
higher genus Riemann surface. Experimental limitations are thus an
obstacle to analyse the effects of non-trivial topologies on quantum
systems.

The reunion of these two topics, namely quantum simulation and
topology, is a natural and tantalizing evolution for both sets of
ideas. So far, the focus in quantum simulation has been on topological
properties emerging in infinite systems due to their dynamics, e.g.,
in the toric code \cite{kitaev2003fault, weimer2010rydberg,
  barreiro2011open} or in periodically driven systems
\cite{kitagawa2010topological, kitagawa2012observation,
asboth2013bulk}, and 
in synthetic quantum Hall \cite{goldman2012detecting, hugel2014chiral, celi2014synthetic,atala2014observation} 
and quantum spin-Hall 
\cite{goldman2010realistic, hauke2012non, beeler2013spin,jotzu2014experimental, struck2014spin} 
systems that exhibit edge states when subjected to open boundary conditions.
  The search for edge
states includes also theoretical and experimental efforts in
understanding Majorana fermions, since they are produced at the
boundaries of some quantum systems \cite{fu2008superconducting,
  mourik2012signatures, jiang2011majorana}. But, so far, the
simulation of systems with non-trivial boundary conditions, with the
exception of circle/torus geometry (cf. theory
\cite{PhysRevA.79.063616,PhysRevA.62.063610,PhysRevA.62.063611}, and
experiments
\cite{ramanathan2011superflow,eckel2014direct,PhysRevLett.113.045305,
  PhysRevA.88.063633,PhysRevA.88.053615,PhysRevLett.110.025302,PhysRevLett.106.130401},
and references therein), has been very scantily explored (see also
\cite{beugeling2014nontrivial,quelle2014topological}, which appeared
while this work was in progress).

Geometry and topology have already made their appearance in quantum
simulators, specifically in optical lattices with ultracold
atoms. Recently, there have been proposals to simulate quantum
many-body physics in certain types of curved background spacetimes
\cite{boada2011dirac}, tailoring the hopping amplitudes of the optical
lattice. Moreover, in \cite{boada2012quantum} a protocol was
introduced to use the different atomic states as an artificial extra
dimension. This latter proposal plays a key role in our approach to
the simulation of quantum matter in different topologies. In effect,
by managing the internal interactions between the internal states, we
will show how to turn an open 1D optical lattice into a system with
periodic boundary conditions, a cylinder, a torus or a M\"obius strip.
Our proposal can be engineered also using other platforms and/or may be
combined with 
other techniques  
such as the ones allowing for well-established toroidal compactifications 
\cite{PhysRevLett.99.260401,0953-4075-34-4-105,PhysRevA.89.053602,PhysRevA.87.013619,
  PhysRevA.87.043601,PhysRevA.81.043619,PhysRevA.79.043620,stabilityKavoulakis,
  PhysRevA.74.065601,PhysRevA.75.063406,PhysRevA.72.063612,0953-4075-37-22-001,PhysRevA.66.033602,PhysRevA.63.013608,PhysRevA.59.2990,0295-5075-46-3-275}, 
or the speckle potentials allowing to simulate in a controlled way disorder \cite{bakr2009quantum, sherson2010single}.

The paper is organized as follows. Section
\ref{sec:artificial_topology} presents the general strategy to
simulate non-trivial topology on a quantum system, while the
experimental aspects are discussed in section \ref{sec:imple}. Section
\ref{sec:signatures} is devoted to an analysis of signatures of
non-trivial topological effects, which can be observed in systems
amenable to experimental realization. We end up, in section
\ref{sec:conclusions} presenting the conclusions and a discussion of
the possibilities for future work.


\section{Artificial topology}
\label{sec:artificial_topology}

The general aim of our work is to build quantum simulators for
dynamics in different topologies out of an optical lattice, which
naturally have open boundary conditions. In order to illustrate our
strategy let us start with the simplest paradigmatic example:
simulating quantum dynamics on a ring i.e., a 1D system with periodic
boundary conditions (PBC). In principle, this can be achieved by
embedding it into a plane, bending it into a circumference and
creating an effective interaction between the two extremes which is
identical to the one in the bulk. Thus, an extra dimension is
required, as well as the possibility of bending the system without
altering its dynamical properties. Both requirements are difficult to
meet. Therefore, we shall explore a different possibility, which
amounts to engineering an artificial extra dimension.

\begin{figure}
\begin{centering}
\begin{tabular}{c}
\includegraphics[width=.7\columnwidth]{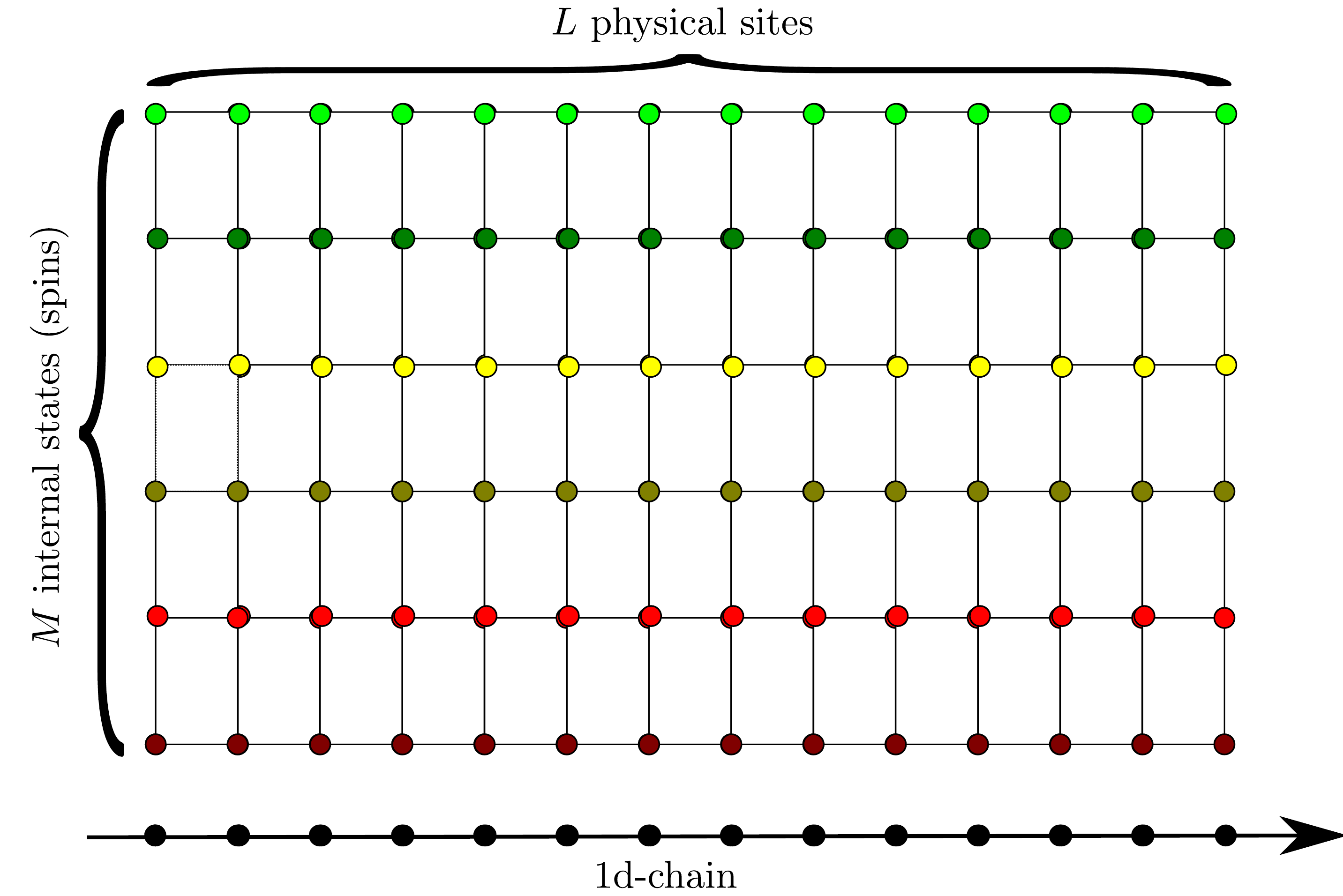} \tabularnewline
\end{tabular}
\par\end{centering}
\caption{\label{fig:synthetic_lattice} Idea of a synthetic lattice: a
  1D chain of length $L$ sites with $M$ species is equivalent to a
  $L\times M$ synthetic lattice, once the chain is dressed with
  appropriate couplings between species.}
\end{figure}

For definiteness, let us discuss a bosonic 1D hopping model with $L$
sites whose Hamiltonian is merely kinetic. Let $a^\dagger_i$ create a
boson at site $i$. The PBC are obtained by connecting the end points
with an extra term.

\begin{equation}
  H_c= -\( J \sum_{i=1}^{L-1} a_i^\dagger a_{i+1} 
+ J_c a_1^\dagger a_L \) + H.c.,
\label{eq:pbc}
\end{equation}
where $J_c$, the closing hopping, should be taken as $J_c=J$. The
problem of simulating an $S^1$ topology is tantamount to generating
this closing term which connects both boundaries of the system. There
are several generic strategies to create that term:

\begin{itemize}
\item Embed the system in a plane and bend it until both boundaries
  touch, thus reducing the boundary term to an ordinary bulk term.
\item Induce a long-range hopping through a medium or an intermediate
  state.
\item Use a synthetic dimension.
\end{itemize}
This work focuses on the last solution. The introduction of an extra
dimension through internal degrees of freedom was proposed in
\cite{boada2012quantum}. Indeed, an open 1D line of $L$ sites, each
endowed with $M$ internal states, can be regarded as an $L\times M$
synthetic 2D lattice, see Fig. \ref{fig:synthetic_lattice}. In
geometric terms, we can think of the internal states as a fiber
opening at each real-space site. The resulting synthetic lattice would
be, therefore, a discrete analogue of a fibre bundle. A generic
hopping Hamiltonian for this system can be written as

\begin{equation}
H = -\( \sum_{\sigma,\sigma'} \sum_{i=1}^{L-1} J^v_{i,\sigma,\sigma'}
b^{\dagger(\sigma)}_i b^{(\sigma')}_i + J^h_{i,\sigma,\sigma'}
b^{\dagger(\sigma)}_i b^{(\sigma')}_{i+1} \) + H.c.,
\label{eq:generic_1d}
\end{equation}
where $J^v$ and $J^h$ are sets of vertical and horizontal hoppings, in
the synthetic lattice view. The vertical term allows us to connect any
pair of internal states in the same physical site, while the
horizontal term allows us to connect any two internal states in
physically neighboring sites.

\begin{figure}
\begin{centering}
\begin{tabular}{c}
\includegraphics[width=.7\columnwidth]{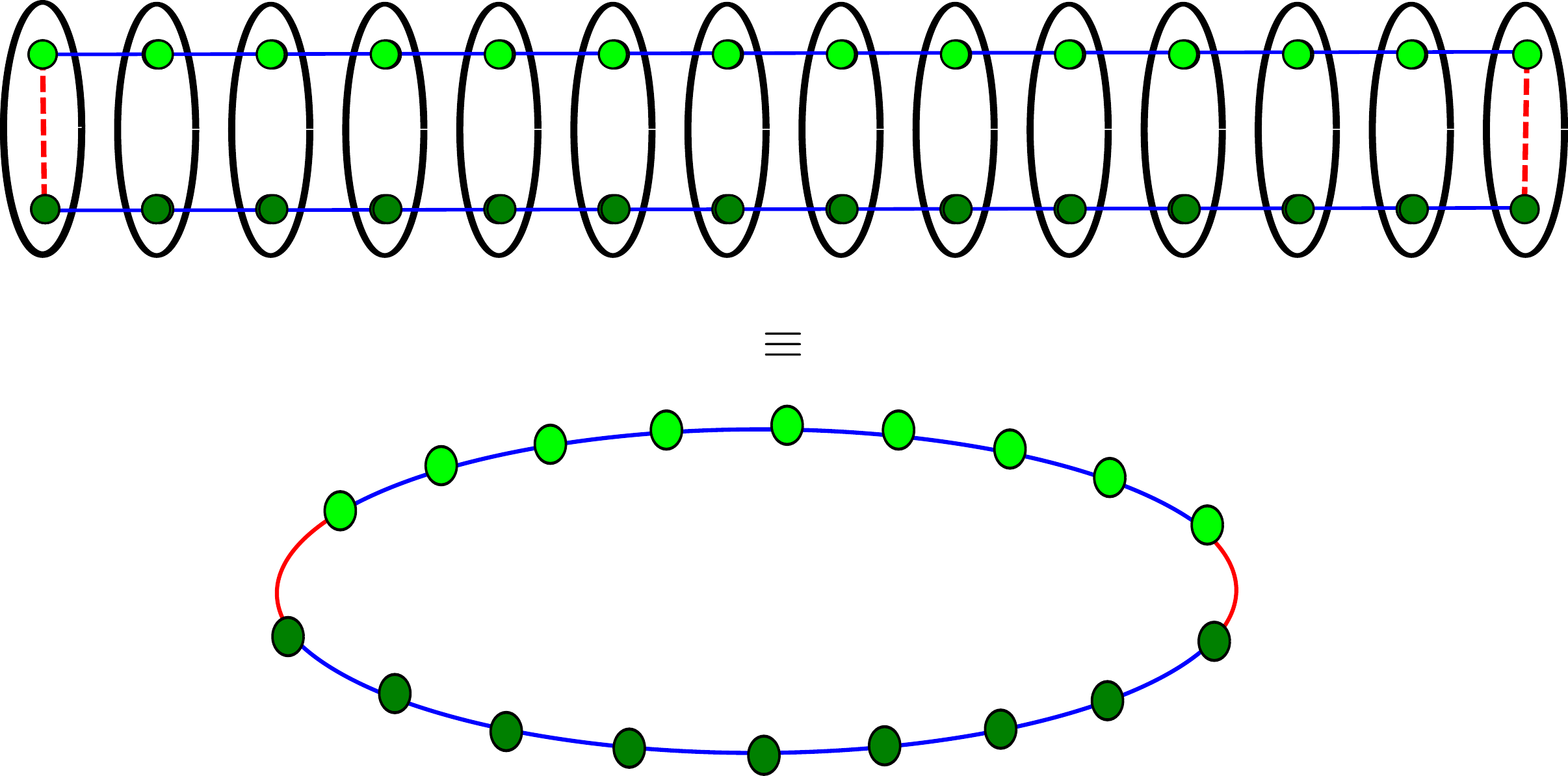} \tabularnewline
\end{tabular}
\par\end{centering}
\caption{\label{fig:circular} Engineering a circumference of $2L$
  lattice sites from a $L$ synthetic lattice that carries $M=2$
  species. The blue solid links indicate the hopping along the chain,
  e.g., the free hopping in a cold-atom implementation, while the two
  dashed red links are induced local interactions between the two
  species, e.g., a Raman coupling between hyperfine levels of
  atoms. The vertical black ellipses represent the physical sites in
  the real 1D-chain occupied by the two species.}
\end{figure}

Figure \ref{fig:circular} illustrates the process by which we can convert an
open spin chain with $M=2$ states per site into a system with PBC. In
Eq. \eqref{eq:generic_1d}, simply set
$J^h_{i,\sigma,\sigma'}=J\;\delta_{\sigma,\sigma'}$ and
$J^v_{i,\sigma,\sigma'}$ to be zero in the bulk, but not in the
extremes, $i=1$ or $i=L$, in which case we have a connecting term
between the two species: $J^v_{i,1,2}=J$. If we introduce $2L$ virtual
particle creation operators, $a_j=b^{(1)}_j$ and
$a_{2L+1-j}=b^{(2)}_j$, $j=1,\dots,L$,  the Hamiltonian reads exactly as a 1D-PBC
hopping Hamiltonian. 

Let us summarize the idea. By inducing appropriate hoppings on the
internal degrees of freedom ---whether we call them species, spin
values, etc.--- we can attain effectively higher-dimensional dynamics,
giving them a geometric meaning \cite{boada2012quantum}. This higher
dimension can be bent and sewn in different ways, as shown in the
previous example. The simplest application consists on turning an open
chain with two species into a closed one with a single species. It
only needs localized control of the transformations between the
species at the boundaries of the open system.

As an additional feature, our synthetic approach allows to control the
phases of the induced hoppings. This is equivalent to inducing a
magnetic field piercing the chain and, via a gauge transformation, to
create boundary conditions which interpolate continuously between
periodic and anti-periodic ones. In critical 1D spin models a
non-trivial magnetic flux can be regarded as a defect in the
associated conformal field theory (CFT)
\cite{alcaraz1987conformal,alcaraz1988conformal}.

The 1D-PBC lattice described above is the basic building block for
more interesting 2D models. In the next sections we will discuss more
exotic boundary conditions, such as M\"obius strips.

\subsection{Assembling cylinder and torus}\label{sec:cyl_tor}

\begin{figure}
\begin{centering}
\begin{tabular}{c}
\includegraphics[width=.98\columnwidth]{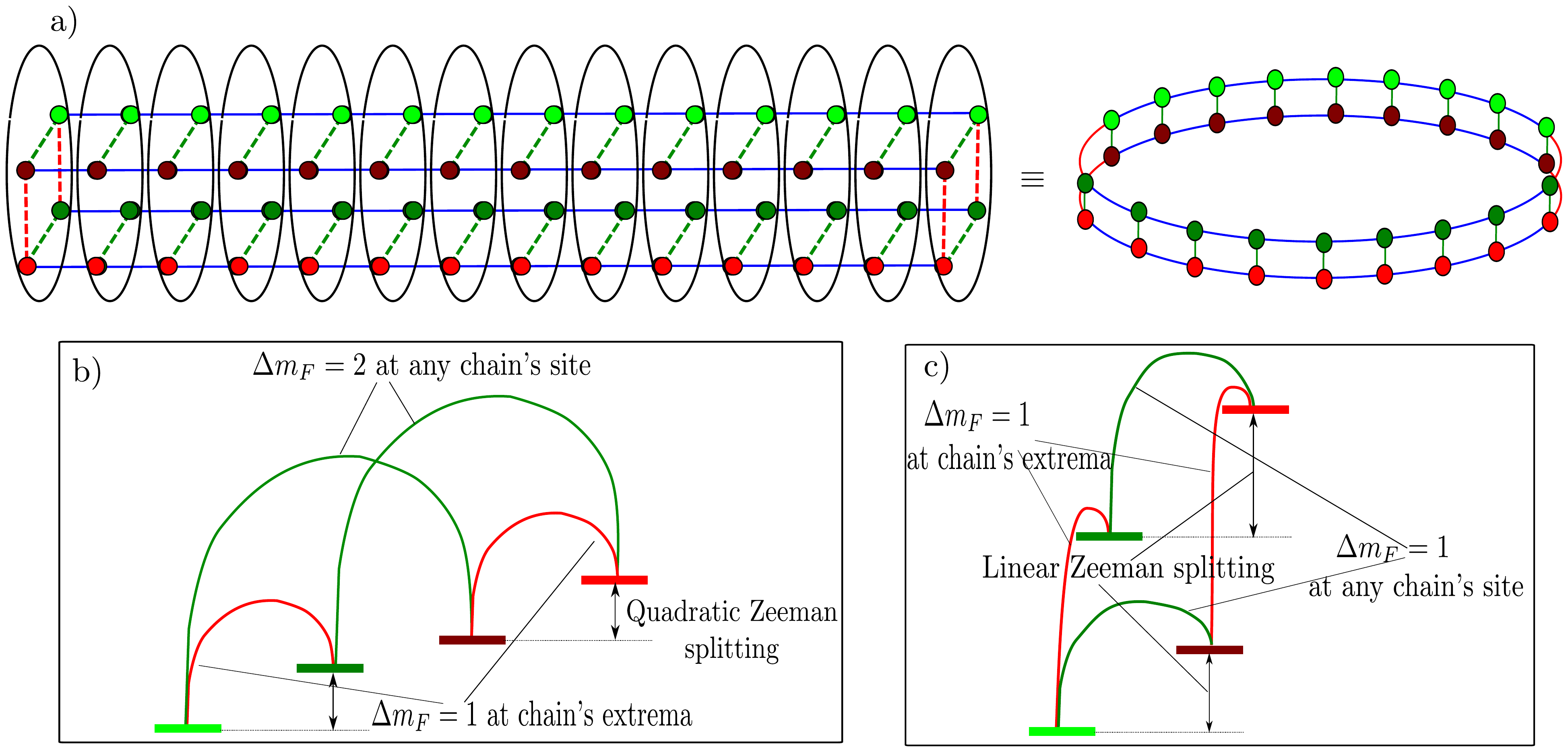} \tabularnewline
\end{tabular}
\par\end{centering}
\caption{\label{fig:cylinder} Engineering a cylinder of basis $L_x$
  and height $L_y$ lattice sites from a $L\times M$ synthetic lattice
  with $L=L_x/2$ and $M=2L_y$. (a) $L_y=2$ synthetic cylinder. In a
  cold atom implementation, the blue solid links indicate the free
  hopping while the dashed ones are laser or radio-frequency
  induced. The dashed red links are the ``sewing'' hoppings
  $C_{\sigma\sigma'}$ closing each circle of the cylinder, while the
  dashed green ones are the hoppings $B_{\sigma\sigma'}$ along its
  height, connecting the different circles. (b) and (c)
  Spectroscopical arrangement of the four internal degrees of freedom
  in the one-manifold and two-manifold scheme, respectively. The
  former can be realized by using the groundstate of atoms $F\ge \frac
  32$ like Li, K, Yb, Sr, Er, etc, and requires quadratic Zeeman
  splitting in order to have $J'$-coupling (in red) only between
  odd-even $m_F$-states. The latter requires earth-alkali like atoms
  with $F\ge \frac 12$ as $^{171}{}$Yb, $^{173}{}$Yb, and $^{87}{}$Sr
  and in this case the linear splitting is sufficient in order to make
  the spin accessible by Raman lasers or radio-frequency pulses. }
\end{figure}

A cylinder can be understood as a fiber bundle of segments emerging
from each point of a circumference. Let us describe how to create
cylindricalal synthetic lattice (i.e., a ladder with PBC) of size
$2L_x\times L_y$ from a 1D open chain of $L_x$ real sites, with
$M=2L_y$ internal states per site. Let $a^\dagger_{i,j}$ create a
particle at site $(i,j)$ of the cylinder. Its correspondent in the
synthetic lattice will be $b^{\dagger(2j-1)}_i$ if $i\leq L_x$ and
$b^{\dagger(2j)}_{2L_x+1-i}$ otherwise. An example with $L_y=2$ is
shown in figure \ref{fig:cylinder}.

The Hamiltonian of a free bosonic system on a cylinder can be written as
\begin{align}
H=&-J\( \sum_{i=1}^{2L_x}\sum_{j=1}^{L_y}  a^\dagger_{i,j} (
a_{i+1,j}+a_{i,j+1} ) \right. \nonumber \\
&\left.+\sum_{j=1}^{L_y} a^\dagger_{2L_x,j} a_{1,j}\) + H.c.
\label{eq:ham_cylinder}
\end{align}
which can be mapped to the form \eqref{eq:generic_1d}. If $i\neq 1$
and $i\neq L$, we have

\begin{equation}
\begin{array}{l l}
J^h_{i,\sigma,\sigma'}&=J\;\delta_{\sigma,\sigma'} \\
J^v_{i,\sigma,\sigma'}&=J\; B_{\sigma,\sigma'} \\
\end{array}
\label{eq:cylinder_bulk}
\end{equation}
where $B_{\sigma,\sigma'}$ is a $2L_y\times 2L_y$ bulk Hermitian
matrix of internal hoppings implementing motion in the transverse
direction of the cylinder

\begin{equation}
B_{\sigma,\sigma'}=\delta_{\sigma,\sigma'\pm 2},
\label{eq:bulk_matrix}
\end{equation}
i.e., it is always possible to jump between internal states differing
by two units. On the extremes, for $i=1$ or $i=L$, we have to add a
new term

\begin{equation}
J^v_{i,\sigma,\sigma'}=J\; B_{\sigma,\sigma'} 
+ J\; C_{\sigma,\sigma'} 
\label{eq:cylinder_border}
\end{equation}
where $C_{\sigma,\sigma'}$ is a closing Hermitian matrix which is
responsible for sewing the open edges of the cylinder. Since it
corresponds to a pile of circumferences, the non-zero entries of those
matrix are of the form $C_{2j-1,2j}$ and $C_{2j,2j-1}$. The
geometrical meaning of that closing matrix is that each horizontal
line bends on itself, without mixing.
 
\begin{figure}
\begin{centering}
\begin{tabular}{c}
\includegraphics[width=.98\columnwidth]{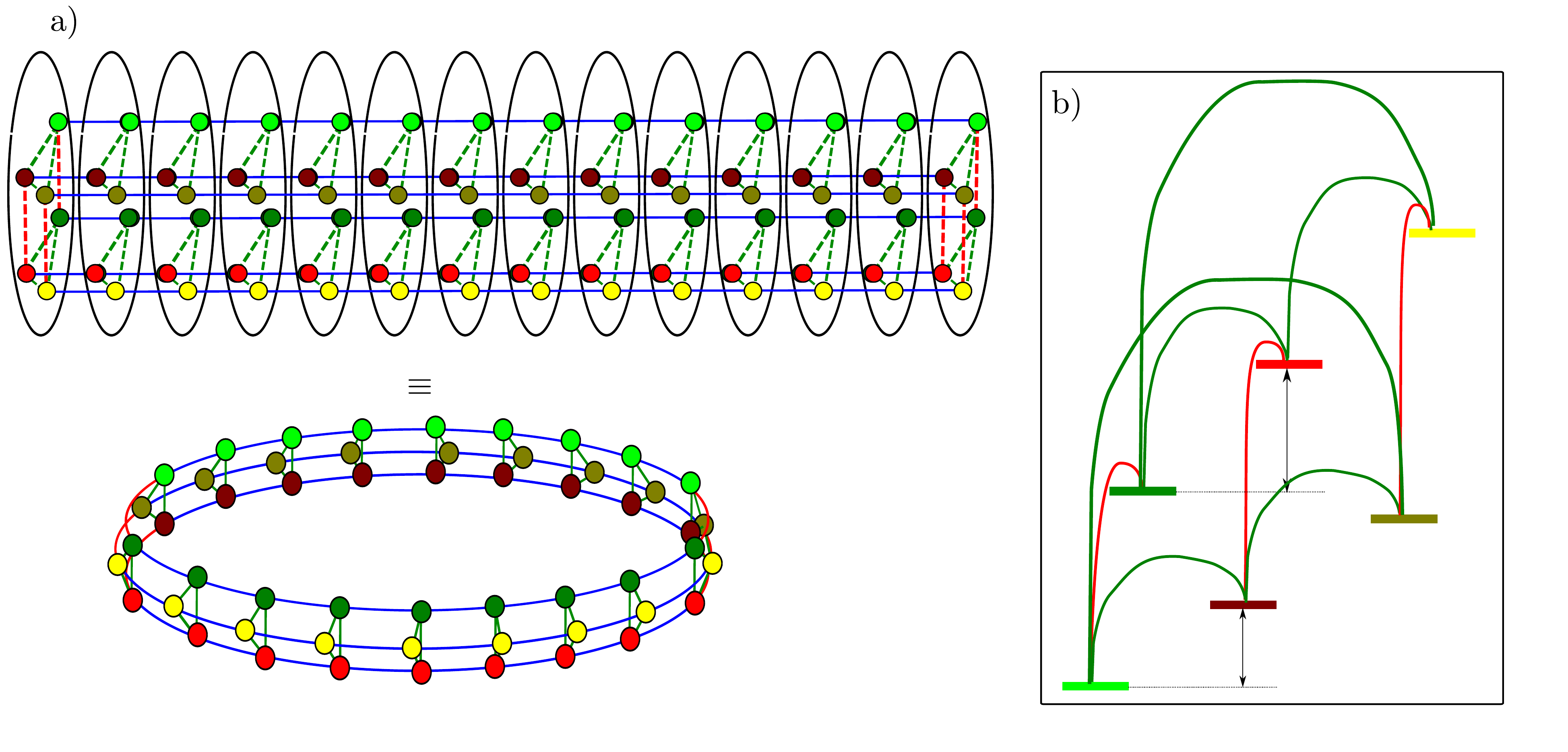} \tabularnewline
\end{tabular}
\par\end{centering}
\caption{\label{fig:torus} Engineering a torus by converting each
  fiber into a circumference. In (a) we show the $L_y=3$ synthetic
  torus.  In a cold atom implementation, the blue solid links indicate
  the free hopping while the dashed ones are laser or radio-frequency
  induced. The dashed red links are unchanged with respect to
  Fig. \ref{fig:cylinder} (a), while additional green dashes ones are
  used to glue the basis of the cylinder connecting the top and bottom
  circles, and correspond to the additional terms in
  $B_{\sigma\sigma'}$. In (b) we show the spectroscopical arrangement
  of the six internal degrees of freedom in the two-manifold scheme
  requiring $F\ge \frac 32$, for instance, $^{173}{}$Yb.}
\end{figure}

The synthetic cylinder we just described can be easily turned into a
torus by changing matrix $B_{\sigma,\sigma'}$, with the introduction
of new non-zero terms
$B_{1,2L_y-1}=B_{2L_y-1,1}=B_{2,2L_y}=B_{2L_y,2}=1$ which sew together
the upper and lower ends of each fiber, see Fig. \ref{fig:torus}. This
construction makes sense for $L_y\ge 3$.  In other terms, each fiber
becomes a circumference instead of an open segment.  However, while
the number of layers, $L_y$, of the cylinder are limited only by the
total number of internal species available, in the case of the torus
$L_y$ can be further restricted by the ability of coupling the
internal species cyclically (see sect. \ref{sec:imple} for cold atom
implementation).

\subsection{M\"obius strip and twisted torus}\label{sec:moe}

The analogy of the synthetic lattice and the fiber bundle, with the
internal states playing the role of the fiber, can be exploited
further. We can glue the fibers opening at different sites in a
different way, in order to provide a non-trivial topology to the
manifold. For example, by gluing the first and last fibers of a
cylinder via a reflection we can turn it into a non-orientable
manifold, a M\"obius strip.

\begin{figure}
\begin{centering}
\begin{tabular}{c}
\includegraphics[width=.98\columnwidth]{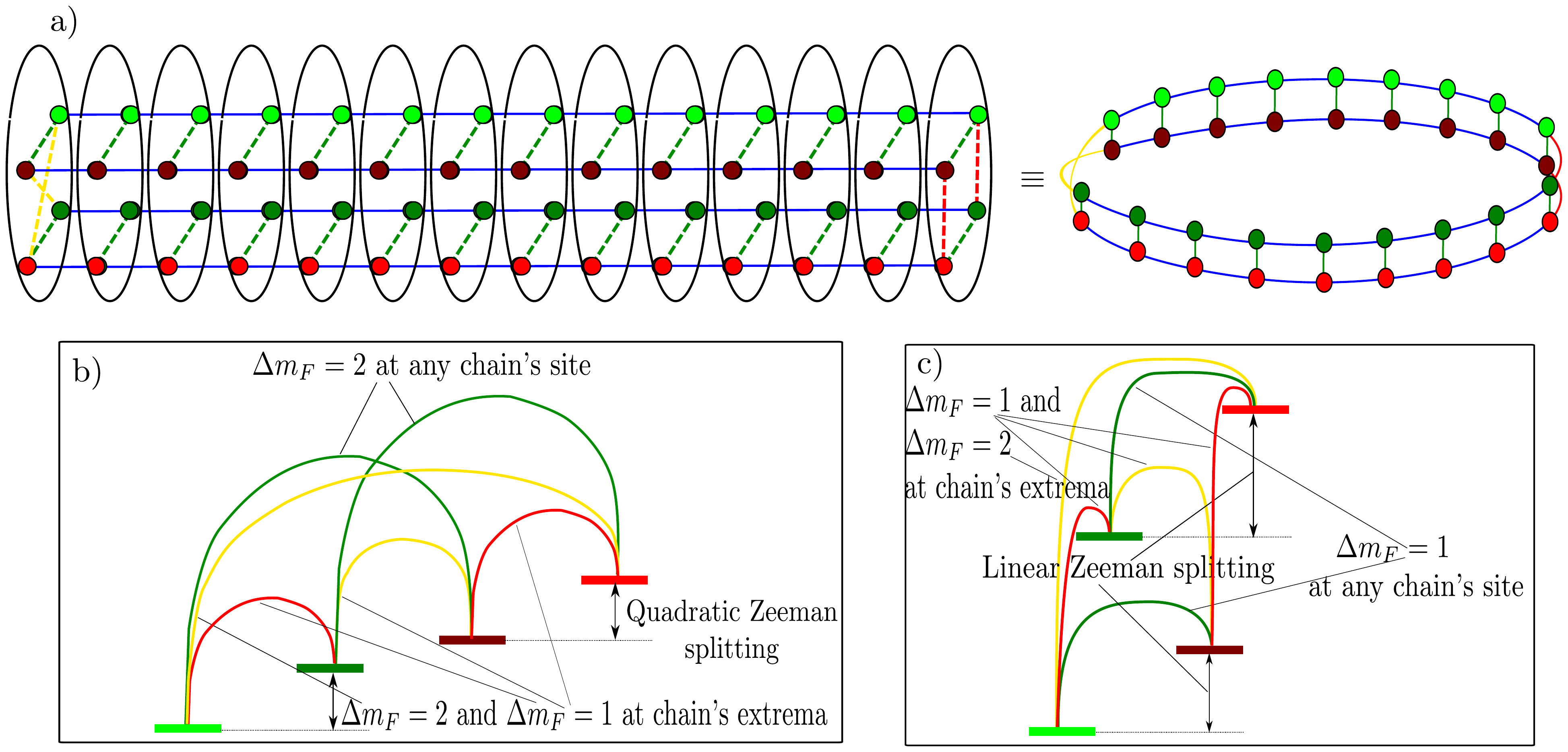} \tabularnewline
\end{tabular}
\par\end{centering}
\caption{\label{fig:Mobius} Engineering a M\"obius strip by twisting
  the cylinder. In (a) we show the $L_y=2$ synthetic strip. In a cold
  atom implementation, the blue solid links indicate the free hopping
  while the dashed ones are laser or radio-frequency induced. The
  dashed green links and half of the dashed red links are unchanged
  with respect to Fig. \ref{fig:cylinder} (a), while the yellow dashed
  ones are the twisted closing hoppings connecting different circles,
  reflecting the change from $C_{\sigma\sigma'}$ to
  $C^M_{\sigma\sigma'}$. In (b) and (c) we see the spectroscopical
  arrangement of the four internal degrees of freedom in the
  one-manifold and two-manifold scheme, respectively. The former can
  be realized by using the ground state of atoms $F\ge \frac 32$ like
  Li, K, Yb, Sr, Er, etc., while the latter requires earth-alkali like
  atoms with $F\ge \frac 12$ as $^{171}{}$Yb, $^{173}{}$Yb, and
  $^{87}{}$Sr. in both cases, the hyperfine levels are linearly split
  in order to make them accessible by multiple Raman lasers or
  radio-frequency pulses.}
\end{figure}

Let us discuss in detail the construction of the artificial M\"obius
strip, a $2L_x\times L_y$ ladder with twisted boundary conditions,
i.e., site $(2L_x,j)$ is connected to site $(1,L_y+1-j)$. The free
Hamiltonian  reads
\begin{align}
H=&J\( \sum_{i=1}^{2L_x-1}\sum_{j=1}^{L_y-1} a^\dagger_{i,j}
(a_{i+1,j}+a_{i,j+1}) \right. \nonumber\\
&\left.+  \sum_{j=1}^{L_y-1} a^\dagger_{2L_x,j}a_{1,L_y+1-j}\) + H.c. 
\label{eq:ham_Mobius}
\end{align}
The corresponding synthetic lattice Hamiltonian corresponds to the
general form \eqref{eq:generic_1d}, with the following choice of
hoppings, see Fig. \ref{fig:Mobius}. For sites $i\neq 1$ and $i\neq
L$, they are the same as for the cylinder,
Eq. \eqref{eq:cylinder_bulk}. For the extremes, one of them should be
the same as for the cylinder, say $i=L$. But $i=1$ must be twisted
and connect the different values of the transverse coordinate
\begin{equation}
J^v_{1,\sigma,\sigma'} = J\; B_{\sigma,\sigma'} 
+ J\; C^M_{\sigma,\sigma'}
\label{eq:hoppings_Mobius}
\end{equation}
where $B_{\sigma,\sigma'}$ is the bulk matrix, given by
Eq. \eqref{eq:bulk_matrix} and the M\"obius closing matrix,
$C^M_{\sigma,\sigma'}$ has non-zero terms which revert the site
ordering of the extra dimension, {\sl i.e.}, connects sites $y=j$ with
$y=L_y+1-j$. Therefore, we get that the non-zero elements of $C^M$ have the
form $C^M_{2j-1,2(L_y+1-j)}$ (and symmetric) and
$C^M_{2j,2(L_y+1-j)-1}$. See the $L_y=2$ case exemplified in
Fig. \ref{fig:Mobius}, where at site $i=1$ the (yellow) hoppings glue
the two different circles.

Of course, this scheme presents the handicap that the size of the
transverse direction is not scalable , i.e., it is limited by the
number of internal species available and by our ability to couple
them. But, as we will see in the next sections, already with $L_y=2$
we can obtain substantial differences between the cylinder and the
M\"obius strip.

\begin{figure}
\begin{centering}
\begin{tabular}{c}
\includegraphics[width=.98\columnwidth]{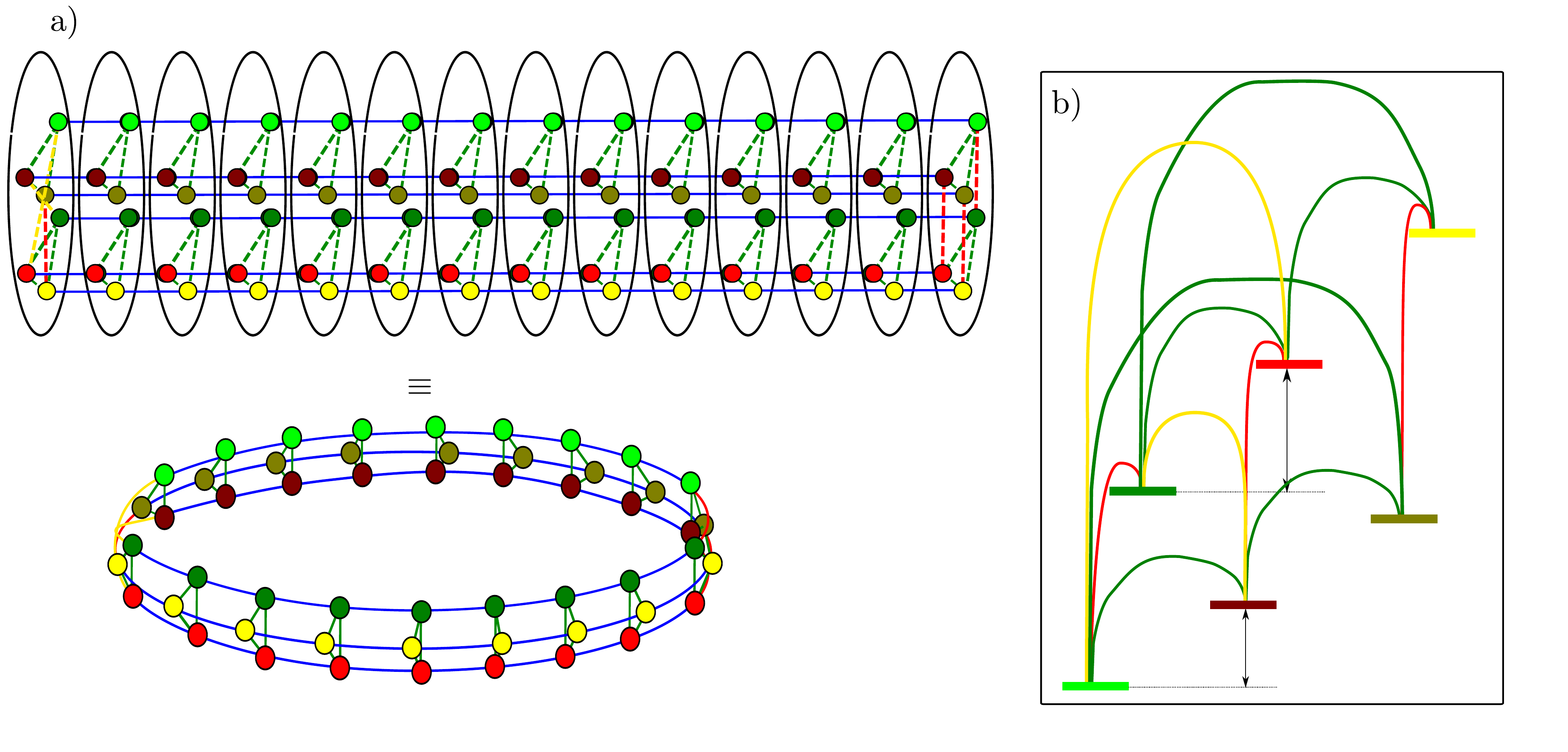} \tabularnewline
\end{tabular}
\par\end{centering}
\caption{\label{fig:twisted_torus} Engineering a twisted torus by
  wrapping the M\"obius strip. Equivalently, a twisted torus can be
  visualized as a torus cut and glued with a twist. In (a) we see the
  $L_y=3$ synthetic twisted torus. In a cold atom implementation, the
  blue solid links indicate the free hopping while the dashed ones are
  laser or radio-frequency induced. The dashed red links are unchanged
  with respect to Fig. \ref{fig:Mobius} (a), while the green dashed
  ones are used to glue the borders of the M\"obius strip. In (b) we
  show the spectroscopical arrangement of the six internal degrees of
  freedom in the two-manifold scheme, which requires $F\ge \frac 32$
  as, for instance, in $^{173}{}$Yb.}
\end{figure}

We can combine the schemes for the M\"obius strip and the torus in
order to build a {\em twisted torus}. The real space Hamiltonian
corresponds to \eqref{eq:ham_Mobius} with the extra term connecting
the $y=1$ and $y=L_y$ values of all fibers: $J\sum_{i=1}^{2L_x}
a^\dagger_{i,1} a_{i,L_y} + H.c.$. This maps into $J\sum_{i=1}^{L_x}
b^{\dagger(1)}_i b^{(2L_y-1)}_i + b^{\dagger(2L_y)}_i b^{(2)}_i$ for
the synthetic lattice Hamiltonian. See Fig. \ref{fig:twisted_torus}
for an illustration.

More general boundary conditions, which do not correspond to a 2D
manifold, are related to the application of a general unitary matrix
of hoppings between sites at $i=2L_x$ and $i=1$
, which can be parametrized as
\begin{equation}
\sum_{j,j'=1}^{L_y} U_{j,j'} a^\dagger_{2L_x,j} a_{1,j'} + H.c.
\label{eq:umatrix}
\end{equation}
where $U\in {\rm U}(L_y)$. If $U$ is the identity matrix, we obtain
the cylinder. Let us now consider the case $L_y=2$. The M\"obius strip
corresponds to the $U=\sigma_x$ case, which has determinant
$-1$. Therefore, can not be connected continuously to the identity
matrix. On the other hand, one can reach a pseudo-M\"obius strip using
a rotation of $\pi$, $U_{1,2}=-U_{2,1}=1$.



\section{Cold Atom Implementation}
\label{sec:imple}

In this section, we show how the artificial topologies described
previously can be made concrete, for instance, in a cold atom set-up.
The basic features that allow to realize the abstract construction of
sect. \ref{sec:artificial_topology} in a cold atom system are the
following:

\begin{itemize}
\item the synthetic lattice is obtained by loading in a 1D
  (spin-independent) optical lattice the atoms, whose hyperfine
  states, belonging to a unique or few hyperfine manifolds, provide
  internal species which form the synthetic dimension;
\item the hopping term $J^h_{i,\sigma,\sigma'}$ of
  \eqref{eq:generic_1d} is the free hopping of atoms in the 1D optical
  lattice and is naturally spin-independent,
  $J^h_{i,\sigma,\sigma'}=J\;\delta_{\sigma,\sigma'}$, as assumed in
  the construction of Hamiltonians \eqref{eq:ham_cylinder} and
  \eqref{eq:ham_Mobius} and their periodic completions;
\item the hopping term $J^v_{i,\sigma,\sigma'}$ is induced and
  tailored by laser and radiofrequency couplings which are local in
  the real-space picture, i.e., are acting on a single site of the 1D
  chain.
\end{itemize}
A similar dictionary can be obtained for other platforms. For
instance, in the spirit of \cite{boada2012quantum}, a synthetic
dimension can be achieved also in photonic crystals as in
\cite{jukic2013four,edge2012metallic}, by changing the connectivity of
the lattice.

It is worth to notice that, while the real spatial dimension is
virtually ``unlimited" (or better scaled up to order $10^2$ lattices
sites), the synthetic dimension is limited always by the number of
atomic internal states available, which is up to 10 for standard atoms
like $^{40}{}$K or $^{87}{}$Sr \cite{stellmer2013production}, but can
be up to 20 in $^{167}{}$Er \cite{aikawa2014reaching}, if just one
hyperfine manifold is taken into account. However, by considering more
than one hyperfine manifold simultaneously or ultracold molecules, see
e.g., \cite{wall2014realizing}, this number can be further increased.
A limited synthetic dimension translates into a limited transverse
dimension of the artificial topology.

Let us start by discussing how to implement the building block of our
construction, a (spinless) periodic chain from a (spinful) open one.
In cold atoms, model \eqref{eq:generic_1d} applied to create PBC can
be realized for instance by loading atoms with at least two hyperfine
(almost degenerate) ground states ($F\ge \frac 12$) in a
spin-independent quasi-1D optical lattice of $L$ sites.  The free
tunneling provides the terms in $J^h$, while the terms in $J^v$ can
be created using Rabi oscillations between the hyperfine states,
induced by Raman lasers focused on sites $1$ and $L$, respectively.
Thus, the synthetic approach we are proposing is essentially local,
since the different species are physically at the same site.  Notice
the {\em scalability} of the procedure: we can build PBC 1D systems of
any size $2L$, if we can build an associated open system with $L$
sites and $M=2$ internal states.

\subsection{Cylinder and Torus}

Let us extend the above construction to the simulation of a cylinder
by layering many circles together as explained in
sect. \ref{sec:cyl_tor}.  It is easy to realize that the Hamiltonian
of \eqref{eq:ham_cylinder} can be implemented with up to two-photon
transitions. Indeed, the most direct arrangement of the internal
degrees of freedom $\sigma$ is in terms of hyperfine states within a
unique hyperfine manifold $F\ge \frac{L_y-1}2$, e.g.,
$\ket{\sigma}=b^{(\sigma)\dagger}\ket{0}=\ket{F,m_F=\bar m+\sigma}$.
For this ordering of the spins, the synthetic sewing coupling
$C_{\sigma\sigma'}$ applied at real-space sites $i=1$ and $i=L$
requires $\Delta m_F=1$, while the synthetic transverse coupling
$B_{\sigma\sigma'}$ requires $\Delta m_F=2$ for any $L_y$. Thus, the
only limitation in $L_y$ is given by the number of available internal
states. It is worth to notice that the coupling $C_{\sigma\sigma'}$
applies only alternatively, i.e., it connects only odd and even spin
values. This implies that the hyperfine states have to be
spectroscopically distinguishable, for instance, through a quadratic
Zeeman splitting.

The spin arrangement considered above is not the only possible one.
Furthermore, two or more (meta)stable hypermanifolds can be
considered. Such a construction is particularly favorable in
earth-alkali like atoms like Yb (see e.g.,
\cite{pagano2014one,cappellini2014direct,scazza2014observation,lahrz2014detecting})
where the optically connected $^1S_0$ and the $^3 P_0$ may be used. In
this case, a convenient arrangement is to place odd (even) $\sigma$'s
in the first (second) manifold, i.e.,
$\ket{2\sigma}=b^{(2\sigma)\dagger} \ket{0}=\ket{I,F,m_F= \bar m +
  \sigma}$, ($\ket{2\sigma+1}=b^{(2\sigma+1)\dagger}
\ket{0}=\ket{II,F,m_F= \bar m + \sigma}$). Thus, the Hamiltonian
\eqref{eq:ham_cylinder} involves in this scheme just $\Delta m_F =1$
transitions.  As further discussed below (cf. sect. \ref{sec:inte}),
the two-manifold construction allows for a richer interaction pattern
than single-manifold one. Both schemes are depicted in
Fig. \ref{fig:cylinder}.

Let us now turn to the implementation of a torus geometry.  As
described in sect. \ref{sec:cyl_tor}, further couplings are needed,
which connect the top and the bottom circles. In the synthetic-lattice
basis, this is equivalent to connecting the last of the odd (even)
spins with the first odd (even) one, for any real-space site. Such a
construction makes sense when $L_y$ is at least 3, the case whose
implementation is detailed in Fig. \ref{fig:torus}.  For simplicity,
let us focus on the two-manifold construction. Here, the additional
coupling requires just $\Delta m_F=2$ and, similarly to the periodic
boundary conditions engineered in \cite{celi2014synthetic}, can be
achieved for instance via a 3-photon transition. For generic $L_y$,
the needed transition has $\Delta m_f=L_y-1$.

\subsection{M\"obius and Twisted Torus}

Let us now discuss the cold atom implementation of a M\"obius
strip. As explained in sect. \ref{sec:moe}, we can get a M\"obius from
the cylinder by replacing the synthetic coupling $C_{\sigma\sigma'}$,
at (e.g.) real-space site $i=1$ with the coupling
$C^M_{\sigma\sigma'}$. This is equivalent to connecting the internal
states $\ket{\sigma=2 l}$ with $\ket{2 (L_y +1 -l)}$, and the states
$\ket{\sigma=2l-1}$ with $\ket{2(L_y-l) +1}$, for $l=1,\dots,L_y$. It
is immediate to realize that for any arrangement of the internal
states as hyperfine states (both in the one- and two-manifold
scenarios) this implies that the maximum $\Delta m_f$ needed to
engineer $C^M_{\sigma\sigma'}$ scales with $L_y$. For instance, in the
two-manifold scheme with the arrangement for the $\sigma$'s described
above, the maximal $\Delta m_F$ is exactly $L_y$, see
Fig. \ref{fig:Mobius} c). Thus, the feasible transverse dimension of
the strip is technically limited, let us say to $L_y=4$, value which
requires at least four-photon transitions.

The step to the implementation of a twisted torus is quite easy and
requires the addition of a coupling with $\Delta m_F= L_y -1$, as
described above.

\subsection{Interactions}\label{sec:inte}

The constructions we have presented produced the kinetic terms of the
Hamiltonians, which are indeed the relevant part for the connectivity
of the model (including boundary conditions) and, thus, for the
hoppings. Cold atom implementation provides a natural way of including
interactions.  In the synthetic-lattice picture, ordinary on-site
interactions due to collisions of atoms with different spins appear as
long-ranged.
   
The pattern of such long-range interactions can be partially
controlled. For instance, it is potentially very different for in the
(i) one-manifold and in the (ii) two-manifold schemes. Interactions
between hyperfine states of the same manifold may change a lot from
atom to atom, but the non-spin-changing ones are in general all of the
same order of magnitude and, for earth-alkali atoms, they are
equal. The spin-changing ones are naturally suppressed ($p$-wave) and
can be enhanced without inducing too high three-body losses by using
optical Feshbach resonances \cite{fedichev1996influence,
  theis2004tuning, goyal2010p}. An alternative route is given by
Raman-induced interactions near $s$-wave Feshbach resonance
\cite{williams2012synthetic, wang2012spin, williams2013raman}.
Collisions between atoms in different manifolds are not affected by
such constraint but are in general lossy.
   
Let us start discussing in details the case (i) with the assumption of
SU$(2F+1)$-symmetric interactions $H_I=\frac U2 \sum_{j} \hat n_j(\hat
n_j -1)$, where $\hat n_j=\sum_{\sigma=1,2L'} b^{(\sigma)\dagger}_j
b^{(\sigma)}_j$ is the total occupation on site $j$ of the physical
1D-chain. As the interaction is invariant under reordering of the
spins, the final Hubbard model on the synthetic cylinder and the
M\"obius strip looks the same for any of the arrangements chosen to
represent the $\sigma$ in terms of $m_F$. Indeed, supposing that we
selectively fill only the spin-states needed i.e., $2L_y$, the local
occupation at site $j$ of the chain becomes the sum of local
occupations at sites $r=j$ and $r'=2L_x+1-j$ of the synthetic lattice,
$\hat n_j= \sum_{l=1}^{L_y} (a_{r,l}^\dagger a_{r,l} +a_{r',l}^\dagger
a_{r',l} )$.  Thus, the interactions will be full range in the
transverse direction at fixed $r$ and for pairs $(r, 2L_y+1-r)$, $H_I=
\frac U2 \sum_{r=1}^{2L_x}\left( {\hat N}_r({\hat N}_r-1) + {\hat N}_r
      {\hat N}_{2L_x+1-r}\right) $, where ${\hat N}_r\equiv
      \sum_{l=1}^{L_y} a_{r,l}^\dagger a_{r,l}$.

The situation is quite different in scenario (ii), even under the
assumption that the interactions are SU$(2F+1)$-invariant in each
hyperfine manifold. To be definite let us consider earth-alkali like
atoms and assume that interactions are negligible in each hyperfine
manifold with respect to the inter-manifold ones, which we model to be
just density-density. The final $H_I$-term in the synthetic lattice is
strongly dependent on the chosen spin arrangement. For instance, we
can engineer a model where only the term $\frac U2 \sum_r {\hat N}_r
{\hat N}_{2L_x+1-r} $ appears.


\section{Topology signatures}
\label{sec:signatures}

Let us discuss possible experimental signatures of the topology of the
underlying manifold showing up in quantum many-body dynamics which are
amenable to experimental observation in our synthetic lattices.

\begin{figure}
\begin{centering}
\begin{tabular}{c}
\includegraphics[width=.98\columnwidth]{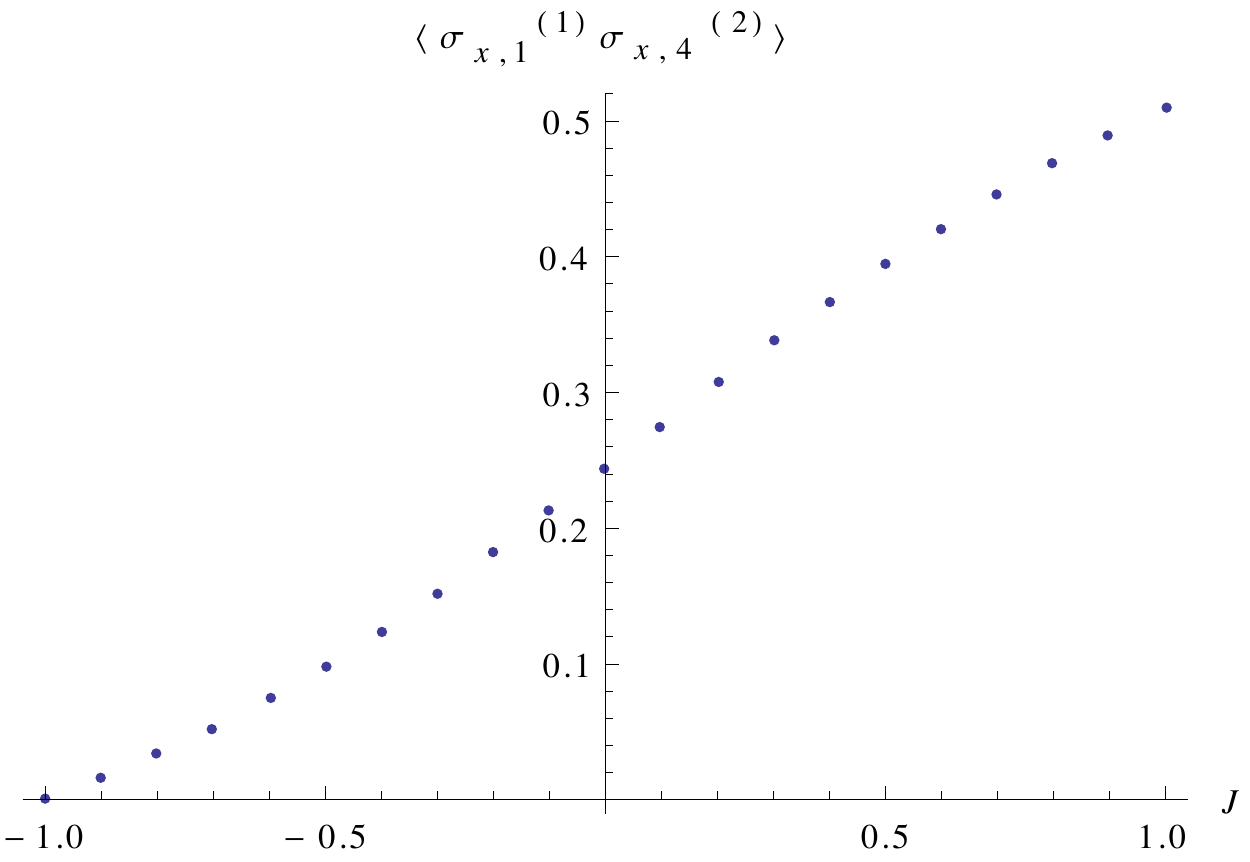} \tabularnewline
\end{tabular}
\par\end{centering}
\caption{\label{fig:ising} Two four-spin Ising chains can be turned
  into a single chain of eight spins by tuning the boundary couplings,
  as shown in Eq. \ref{correlation}. The plot shows the correlation of spins of each species at the different boundaries,
  $B$, as a function of the coupling $J_L \in [-1,1]$, for
  $J_1=\lambda=1$. Note the cancellation of correlations in the case
  of an artificially frustrated boundary. 
      }
\end{figure}

We start by discussing the simplest paradigmatic example of a line
with two species which can be designed to mimic a single species
Hamiltonian on a circumference, as described in Sect. II. In order to
illustrate this idea in a simple way, let us consider the following
Hamiltonian for a double spin chain of length $L$, with tunable
connecting terms at the boundaries:
\begin{eqnarray}
  H=&&\sum_{i=1}^{L-1} \sigma^x_{i,1} \sigma^x_{i+1,1}
    +\lambda \sum_{i=1}^ {L}\sigma^z_{i,1} \nonumber \\ 
    &&+\sum_{i=1}^{L-1} \sigma^x_{i,2} \sigma^x_{i+1,2} 
    +\lambda \sum_{i=1}^ {L}\sigma^z_{i,2} \nonumber \\ 
    &&+J_1 \sigma^x_{1,1} \sigma^x_{1,2}+ 
    J_L \sigma^x_{L,1} \sigma^x_{L,2}.
    \label{spin_exmaple}
\end{eqnarray}
where $\sigma^k_{i,m}$ denotes the $k$-th component of the spin on
site $i$, rung $m$. The two closing interactions $J_1$ and $J_L$ are
responsible for turning the two open chains into a single one on a
circumference. As a consequence, boundary conditions are dictated by
these coefficients. A signature of the artificial boundary conditions
can be measured by the correlation between spins on different chains
at a boundary, namely:
\begin{equation}
  B \equiv \langle \sigma^x_{1,1} \sigma^x_{L,2}\rangle
   -\langle \sigma^x_{1,1} \rangle
   \langle \sigma^x_{L,2} \rangle \ .
\label{correlation}
\end{equation}
Observable $B$ should be zero for disconnected chains. Its value for
fixed $J_1=1$ and $J\equiv J_L \in [-1,1]$ is shown in
Fig. \ref{fig:ising} for $L=4$ and $\lambda=1$. Notice that $J=1$
corresponds to the PBC case, which maximally entangles both chains and
gives the highest correlator. For $J=0$, we obtain a single open BC
chain. More interestingly, for $J=-1$, the twist in the boundary
condition induces a perfect cancellation in the correlator. This
effect is, indeed, a signature of the non-triviality of the
topological effects.

Nonetheless, realistic simulations should model the underlying
geometry by tuning the hoppings of fermions or bosons. We shall now
address such cases, looking for both single-particle and interacting
signatures.

\subsection{Single-Particle signatures}

A natural single-particle playground where we can observe the effect
of the topology is to consider synthetic magnetic fluxes, which boils
down to hoppings with non-trivial phases. In our synthetic lattice it
is very easy to control such phases, in particular to make them
linearly dependent with the position on the chain if the synthetic
links are induced through Raman lasers.

\begin{figure}
\begin{centering}
\begin{tabular}{c}
\includegraphics[width=.7\columnwidth]{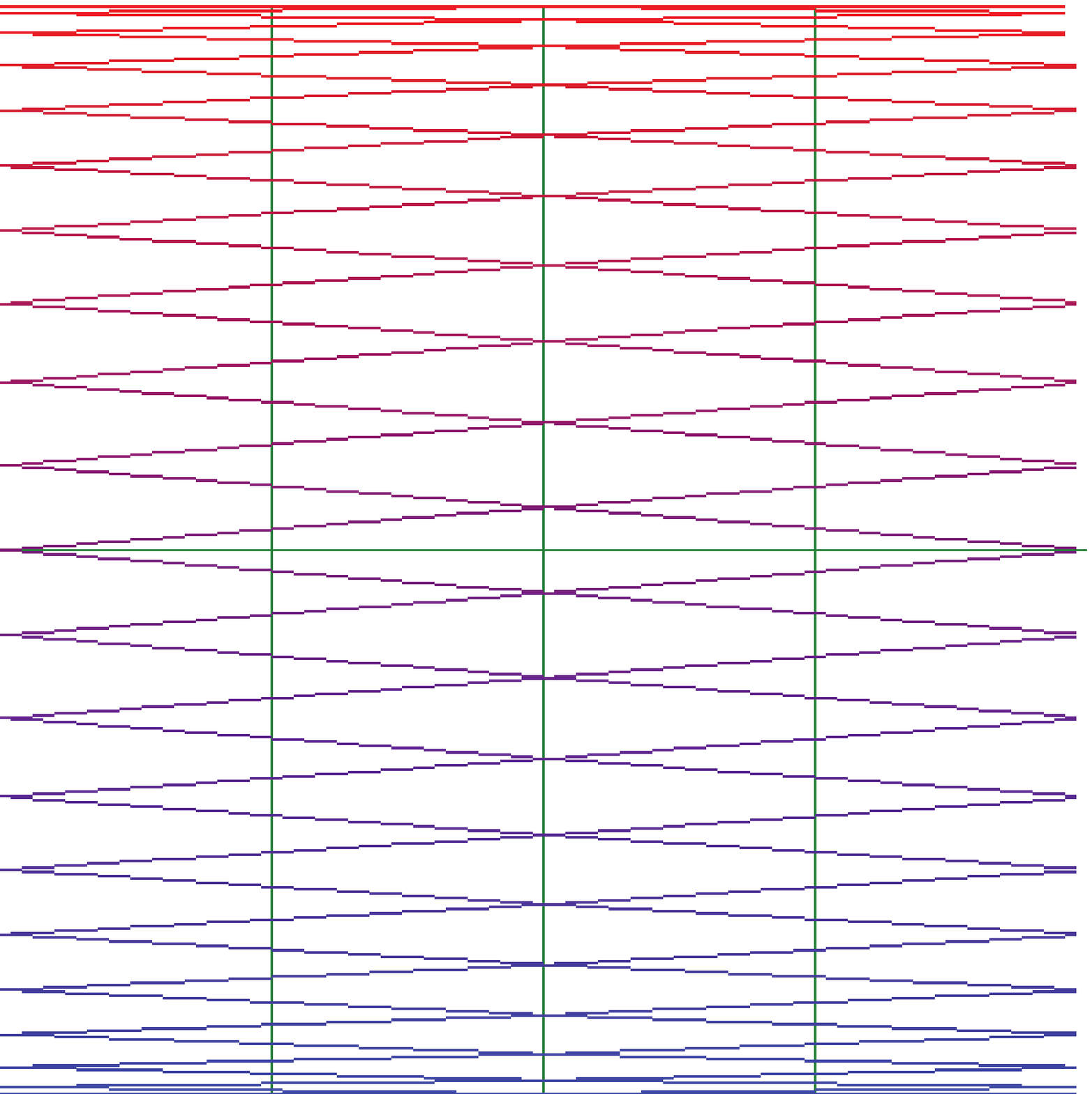} \tabularnewline
\includegraphics[width=.7\columnwidth]{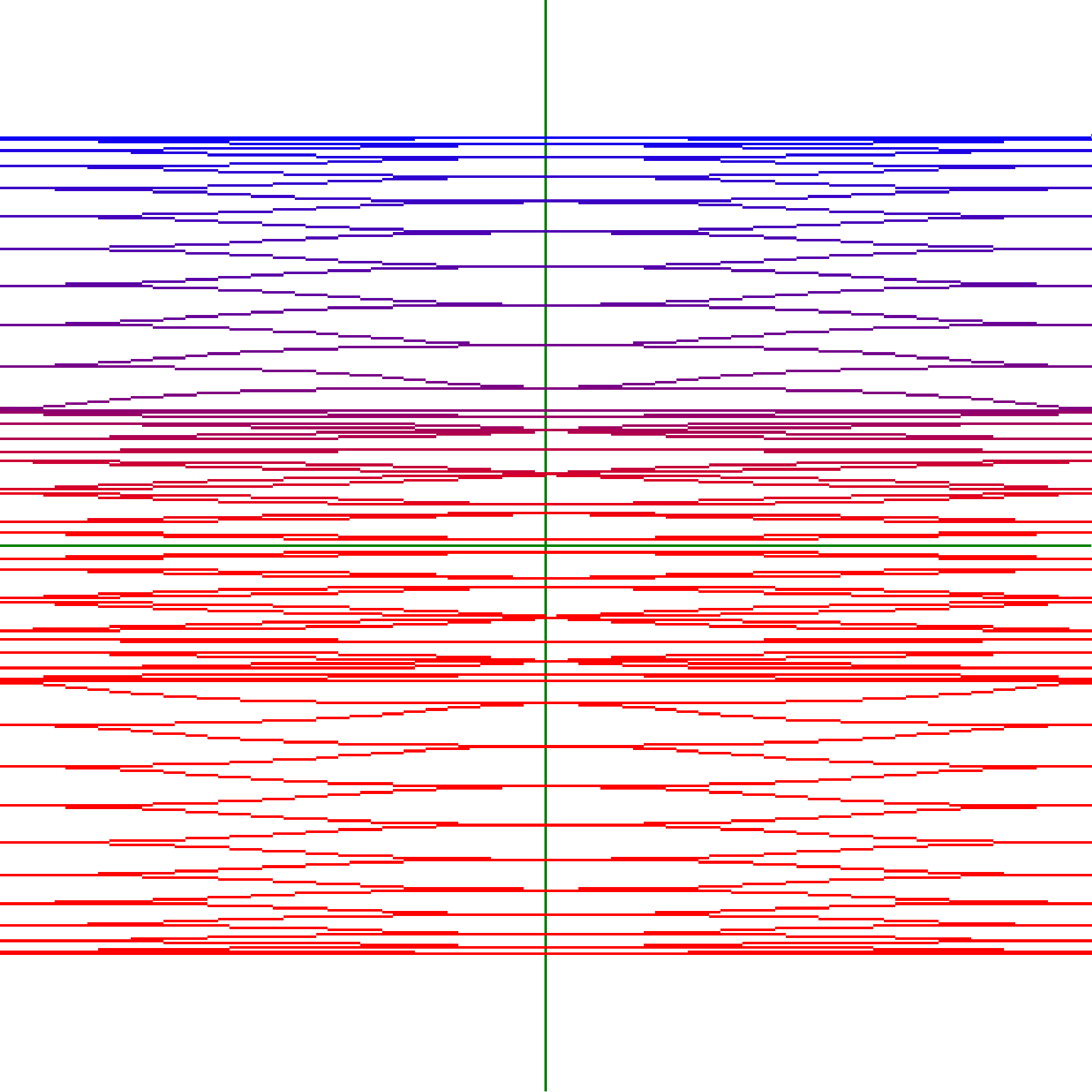} \tabularnewline
\includegraphics[width=.7\columnwidth]{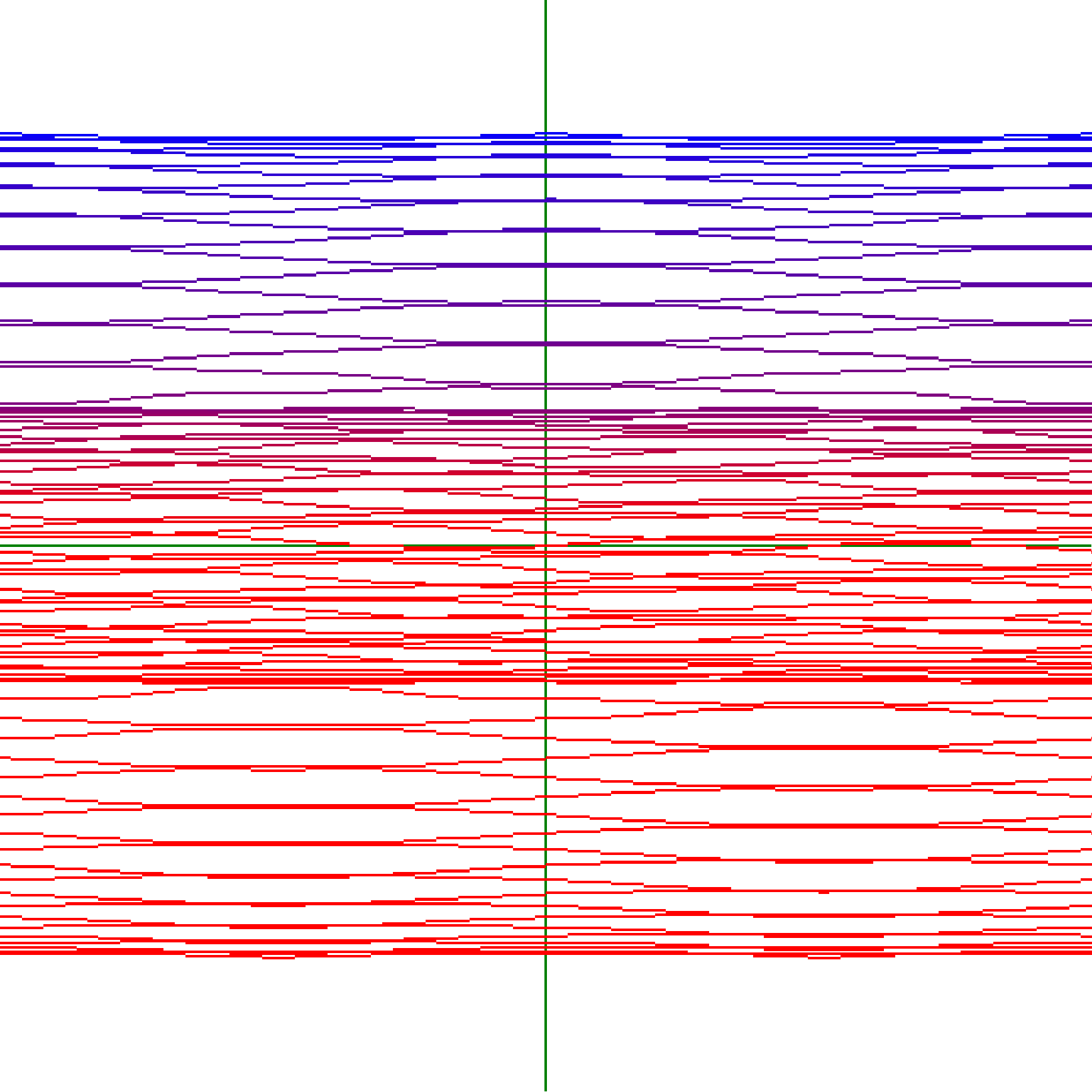} 
\end{tabular}
\par\end{centering}
\caption{\label{fig:hofstadter} Hofstadter-like single-particle
  spectra of several quasi-1D systems under a continuous change of
  boundary conditions. In all cases the $X$-axis is labeled by
  $\phi\in [0,2\pi]$, and color corresponds to eigenvalue index. In
  (a) we present the spectrum for a PBC system such as that in
  Eq. \eqref{eq:pbc} with $L=40$, pierced with a flux $J_c=e^{i\phi}$
  with $\phi\in [0,2\pi]$. In (b) and (c) we show the single-body
  spectrum of a ladder of size $40\times 2$, such as
  Eq. \eqref{eq:ham_cylinder} in which the opposite extremes are
  joined with a unitary matrix of different types: cylindrical in (b)
  and M\"obius-like in (c), as specified in the boundary conditions
  given in equations \eqref{eq:cylinder_umatrix} and
  \eqref{eq:Mobius_umatrix}.}
\end{figure}

Let us start with a 1D Hamiltonian with PBC, as in Eq. \eqref{eq:pbc},
either for spinless fermions or bosons, with an arbitrary closing
phase $J_c=e^{i\phi}$, representing a magnetic flux. Its
single-particle spectrum, as a function of $\phi\in [0,2\pi]$ is shown
in Fig. \ref{fig:hofstadter}. Thus, the left and right extremes are
periodic boundary conditions, while the center corresponds to
anti-periodic ones. Notice that the gap of a fermionic system at
half-filling will evolve continuously, presenting a maximum at
$\phi=\pi$.

The more involved case of a 2-rung ladder is shown in
Fig. \ref{fig:hofstadter} (b) and (c). In those cases we have again a
free Hamiltonian either for spinless fermions or bosons, such as
Eq. \eqref{eq:ham_cylinder} with $L_y=2$. The closing link between the
two extremes can be chosen to be a generic unitary matrix, as shown in
Eq. \eqref{eq:umatrix}. In both cases, we have selected a
one-parameter family of unitary matrices with special properties. In
\ref{fig:hofstadter} (b) it is a rotation of angle $\phi$:

\begin{equation}
U_{+1}(\phi)=
\begin{pmatrix} \cos \phi & -\sin \phi \\ \sin\phi & \cos\phi 
\end{pmatrix}
\label{eq:cylinder_umatrix}
\end{equation}
where the $+1$ stands for the value of the determinant. Thus, for
$\phi=0$ we have the identity matrix, which means cylindrical boundary
conditions. Meanwhile, for (c) we have used a different one-parameter
family:

\begin{equation}
U_{-1}(\phi)=
\begin{pmatrix} \cos\phi & \sin\phi \\ \sin\phi & -\cos\phi
\end{pmatrix}
\label{eq:Mobius_umatrix}
\end{equation}

Although those transformations are unitary, they have determinant
$-1$, and thus can not be connected continuously with the identity
matrix. For $\phi=\pi$ we obtain a M\"obius strip. Notice that the
single-particle spectrum is different in both cases, and thus the
energy gap at half-filling constitute a topological signature.

\begin{figure}
\begin{centering}
\begin{tabular}{c}
\includegraphics[width=.98\columnwidth]{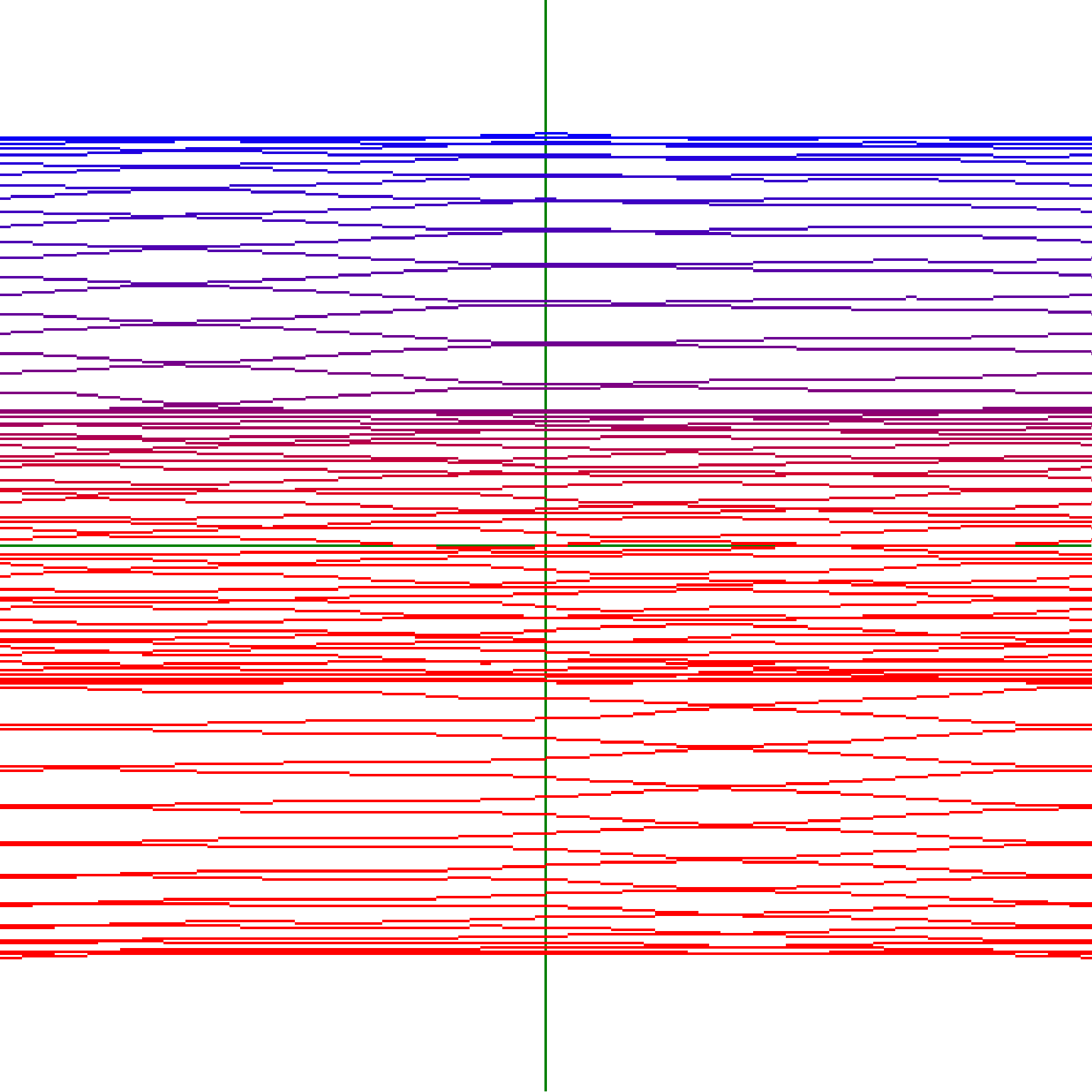} \tabularnewline
\end{tabular}
\par\end{centering}
\caption{\label{fig:cylmoe}Single-particle spectrum of the $40\times
  2$ ladder Hamiltonian represented in Eq. \eqref{eq:ham_cylinder}
  undergoing a smooth transition between a cylinder and a M\"obius
  strip, with boundary conditions specified in
  Eq. \eqref{eq:cylmoe_umatrix}.}
\end{figure}

The magnetic single-particle behavior is sensitive to the
orientability of the underlying lattice. If the lattice is
topologically equivalent to a cylinder and orientable, a constant
magnetic field piercing the surface induces steady counter-propagating
currents on each edge, which are called edge states. Their topological
nature reflects on their robustness under local perturbations. If the
lattice is not orientable, intuition dictates that the current cannot
form as there is no notion of normal to the lattice, i.e., of sign of
the magnetic flux and thus of chirality of the currents. We can check
our intuition by considering the synthetic cylinder and M\"obius strip
as examples of orientable and no-orientable surface with sharp
boundary, respectively. Our synthetic construction allows to smoothly
interpolate between the two by considering a $U$ matrix of the form

\begin{equation}
U_{+1\to -1}(\phi)=
\begin{pmatrix}\cos\phi & -\sin\phi e^{i\phi} \\
  \sin\phi & \cos\phi e^{i\phi}
\end{pmatrix}
\label{eq:cylmoe_umatrix}
\end{equation}
this hopping matrix is also unitary, but its determinant is
$e^{i\phi}$. For $\phi=0$ it is $+1$, and we have the cylinder, while
for $\phi=\pi$ it gets $-1$, and we obtain the M\"obius strip. The
single-particle spectrum of this $L_x\times 2$ ladder with different
boundary conditions is shown in Fig. \ref{fig:cylmoe}.

\subsection{Interacting signatures}

Quantum simulators are not restricted to the study of free
systems. Interactions can typically be tailored to a certain
extent. Our generic Hamiltonian can be written as

\begin{equation}
H= H_K + U\sum_{\<i,j\>} n_in_j - \sum_i \mu_i n_i\, ,
\label{eq:extended_hubbard}
\end{equation}
where $H_K$ is the kinetic Hamiltonian described in
Eq. \eqref{eq:generic_1d}, and $U$ is the strength of the
nearest-neighbor interaction, $\mu_i$ is a local chemical potential
and $n_i$ is the local particle number. The sum in the second term is
over nearest neighbors of a certain adjacency structure, which need
not be the same as the one employed for the kinetic term. We take
$\mu_i$ to be slightly random, in order to remove exact degeneracy
in the ground state. The topology of the underlying lattice are
totally encoded in the kinetic Hamiltonian $H_K$, which is affected by
a global hopping constant $J$.

Let us start by considering a bosonic system with Hamiltonian
\eqref{eq:extended_hubbard} and focus on the local particle-number
fluctuations in the ground state, $\sigma^2=\sum_i
(\<n_i^2\>-\<n_i\>^2)/N$, where $N$ is the total particle number. It
can be employed to distinguish the different phases. Mean-field
calculations cannot distinguish between different topologies, since
they are local in character so, {\em a fortiori}, it will give the
same estimate for $\sigma^2$ for all boundary conditions. Using exact
diagonalization, on the other hand, different topologies can be told
apart by inspecting the behavior of $\sigma^2$ as a function of
$J/U$. For a large $J/U$ the bosons are in a superfluid state with
large particle-number fluctuations, since each particle is delocalized
over the whole lattice. For small $J/U$ the bosons are localized in a
checkerboard pattern and the particle-number fluctuations are small.
\begin{figure}
\begin{centering}
\includegraphics[width=\columnwidth]{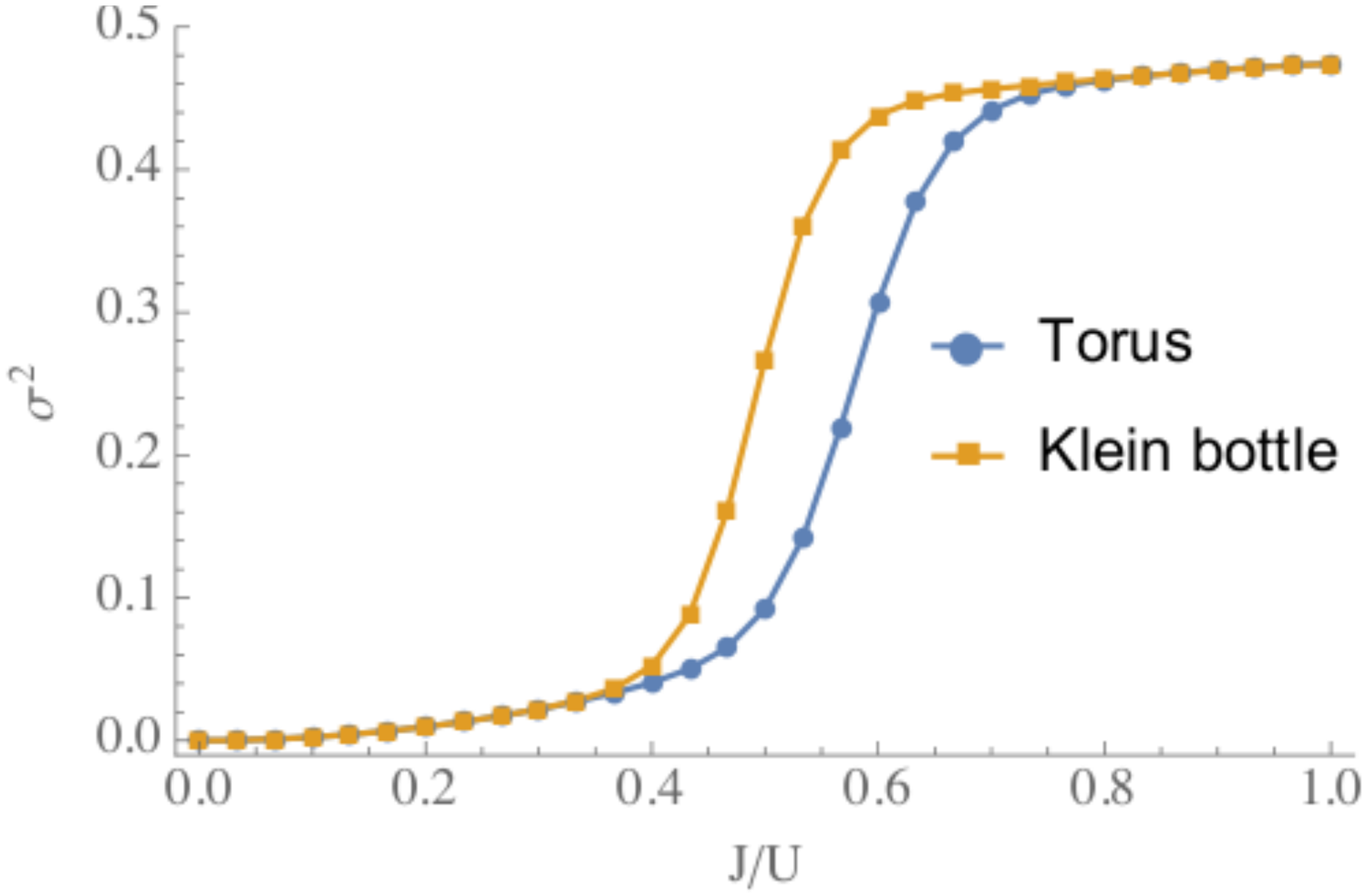}
\vspace{-11\baselineskip}
\caption{\label{fig:hub_klein_fluct}Ground state particle-number
  fluctuations for the Bose-Hubbard model of
  Eq. \ref{eq:extended_hubbard} at half filling, defined on a 2D
  lattice with periodic boundary conditions in both directions --a
  torus-- and on a 2D lattice with the boundary conditions of a Klein
  bottle, computed via exact diagonalization. The results are for a
  4x4 lattice. Similarly to the results for the M\"obius band, the
  ground state in twisted boundary conditions seems to favor larger
  particle-number fluctuations at intermediate values of $J/U$. The
  computation has been done with small disorder in $\mu$ to remove the
  degeneracy at $J=0$, as the two complementary checkerboard coverings
  of the lattice are ground states. 
    }  \par\end{centering}
\end{figure}

\begin{figure}
\begin{centering}
\includegraphics[width=.9\columnwidth]{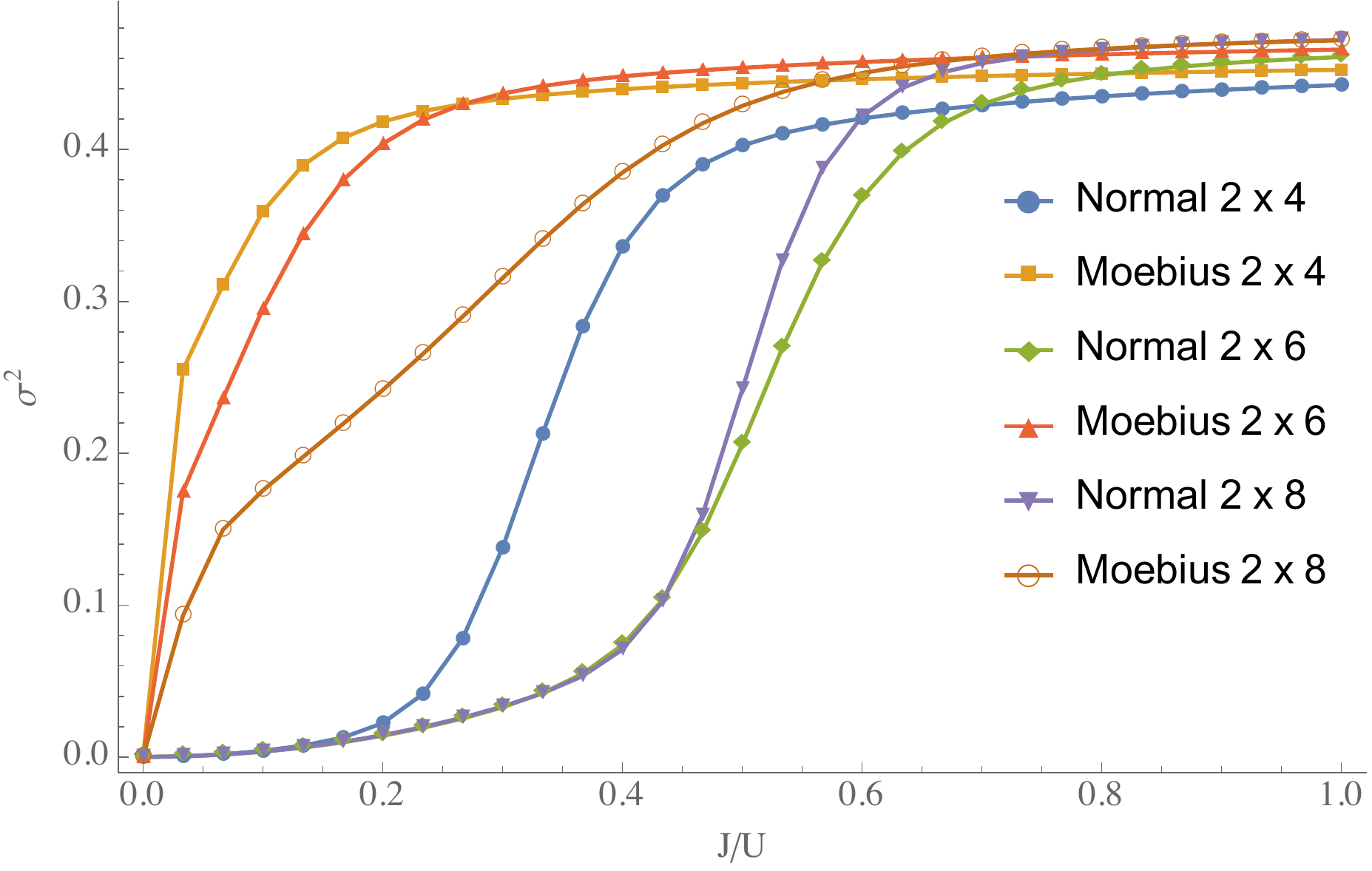}
\caption{\label{hubMobiusfluct} Particle-number fluctuations in the
  ground state of the Bose-Hubbard model of
  Eq. \eqref{eq:extended_hubbard} at half filling defined on a strip
  with periodic and M\"obius boundary conditions computed with exact
  diagonalization. The results are for strips 4, 6 and 8 sites
  long. For large interactions compared to the hopping parameter the
  ground state presents larger particle-number fluctuations for
  twisted boundary conditions than for regular ones. This behavior
  does not disappear as the system size increases, for the range of
  sizes analyzed. The computation has been done with small disorder in
  $\mu$ to remove spurious degeneracy.}  \par\end{centering}
\end{figure}
We consider different boundary conditions for compact lattices (torus
and Klein bottle) and open lattices (cylinder and M\"obius strip) for
different sizes. In Fig. \ref{hubMobiusfluct} we plot $\sigma^2$ as a
function of $J/U$ for the normal strip and the M\"obius strip for
different strip lengths. As expected, in the limits $J/U\rightarrow 0$
and $J/U\rightarrow 1$ the ground state has the same boson number
fluctuations, which is explained by the fact that in both limits the
ground state is a product state in the site basis
\cite{bloch2008many}. In the latter limit this is not exactly the case
due to finite size effects. The data shows that for intermediate
values of $J/U$, where the ground state is entangled, $\sigma^2$ is
sensitive to the different boundary conditions. In
Fig. \ref{fig:hub_klein_fluct} we plot $\sigma^2$ for the ground state
of the Bose-Hubbard model on a torus and on a Klein bottle. The data
shows that $\sigma^2$ can tell the different boundary conditions apart
in this case as well for intermediate values of $J/U$.

Let us now consider a fermionic system with two species per site and a
slightly different dynamics. Let \eqref{eq:extended_hubbard} still be
the Hamiltonian, but we make the repulsion term work only along
vertical lines, i.e., only between particles in the same real-space
site. In terms of the synthetic lattice we can write

\begin{equation}
H = H_K + U \sum_{i=1}^{L_x} n^{(1)}_i n^{(2)}_i + H.c.
\label{eq:fermionic_hubbard}
\end{equation}

For an even $L_x$ we have studied the ground state and first excited
state of Hamiltonian \eqref{eq:fermionic_hubbard} on a cylinder and
M\"obius strip. The first is characterized by the independent motion
of each species. The second by a crossing at the end, where the
species transmute.

\begin{figure}
\begin{centering}
\begin{tabular}{c}
\includegraphics[width=.9\columnwidth]{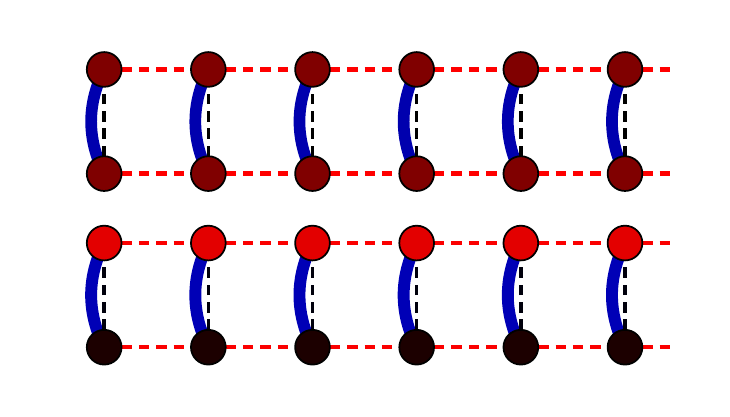} \tabularnewline
\includegraphics[width=.9\columnwidth]{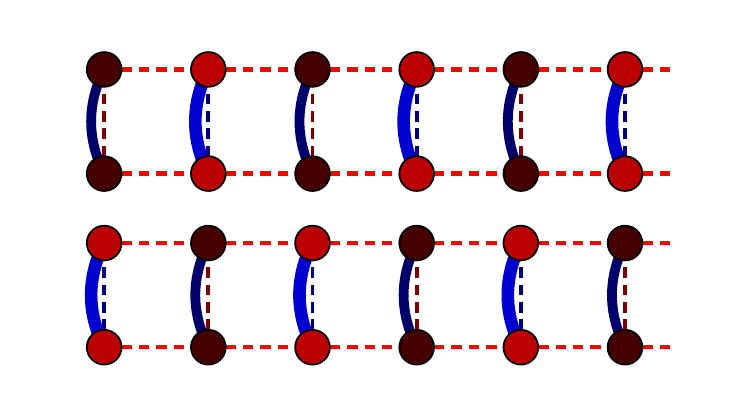} 
\end{tabular}
\par\end{centering}
\caption{\label{fig:fermionic_hubbard} Representation of the ground
  state and first excited state of the fermionic Hubbard model
  \eqref{eq:fermionic_hubbard} on a cylinder and a M\"obius strip of
  size $L_x=6$ and $L_y=2$. The repulsion only takes place along the
  same rung. The color of each node represents the expected value of
  $\<n_i\>$. The color of the bent line represents the correlator
  $\<n_in_j\>-\<n_i\>\<n_j\>$. The dashed lines are the correlators
  $\<a^\dagger_i a_j\>$. }
\end{figure}

Some results are shown in Fig. \ref{fig:fermionic_hubbard}, for
U=J. From top to bottom we see panels (a) and (b), which depict the
ground state and first excited state for the cylinder, and panels (c)
and (d) which show the corresponding M\"obius states. The color of the
circles represent the density, $\<n_i\>$, while the colored arcs
represent the density-density correlator:
$\<n_in_j\>-\<n_i\>\<n_j\>$. The dashed lines represent the hopping
correlator, $\<a^\dagger_i a_j\>$, red being positive and blue
negative in all cases.

The ground state of the cylinder, panel (a) of
\ref{fig:fermionic_hubbard} is characterized by a homogeneous density
and density-density correlators. The first excited state, shown in figure
(b) is doubly degenerate, and it is obtained by adding
one more particle at the upper species. Particles never move between
species, as shown in the null vertical hopping correlators. The
physical picture can be described as follows. The particles move along
their lines in counter-phase, i.e., with highly negative density-density
correlator between the two {\em lines}. This does not lead to
frustration because $L_x$ is even and the lanes never cross.

Panels (c) and (d) show the situation for the M\"obius topology. The
ground state is degenerate, and both states are depicted there. The
local density now shows a checkered pattern, and also the
density-density correlators. The vertical hopping correlators show
also an interesting pattern, alternating positive and negative
values. The physical picture is as follows. The lane crossing induced
by topology makes impossible the previous configuration due to
frustration. The two lanes have become one, and the only possibility
to reduce vertical repulsion is to freeze the system into a
charge-density wave. Particles can not move as fast as they would like
to reduce their kinetic energy, which is an analogue of a {\em traffic
  jam}. That is the reason for the lane changing correlators.

Combining the information of Fig. \ref{fig:fermionic_hubbard} we see
that the Mott transition takes place at different values of the $J/U$
parameter, independently of the system size. This effect is related in
a non-trivial way to frustration.


\section{Conclusions}
\label{sec:conclusions}

We have shown that non-trivial topologies can be simulated by a
combination of two techniques, namely the use of several species at
every spatial degrees of freedom and the generation of couplings among
these species only at the boundaries of the system. In other words,
species work as an extra dimension that allows for the generation of
topological transformations from localized interactions.

In particular we have presented explicit proposals for the realization
of the following geometries:
\begin{itemize}
\item a circle
\item a cylinder
\item a torus
\item a M\"obius strip
\item a twisted torus
\end{itemize}

We have discussed different possibilities of experimental realization
of the proposed schemes, extending significantly the ideas of Ref.
\cite{boada2012quantum}. Finally, we have presented several signatures
of the underlying lattice topology both on free and interacting
systems. These examples involve synthetic gauge fields and synthetic
dimension, including:

\begin{itemize}
\item A two-species open Ising chain with localized interactions among
  them can be converted in a double-length single-species chain with a
  synthetic magnetic field.
\item Hofstadter-like spectra can be obtained for a circle, a cylinder
  and a M\"obius strip.
\item Hubbard systems of moderate size can be engineered on a torus, a
  Klein bottle, a cylinder and a M\"obius strip.
\end{itemize}

Our findings open paths to further investigations of both free and
weakly interacting, as well as strongly correlated systems in optical
lattices with non-trivial topology. Combining such lattice geometries
with synthetic gauge fields leads to various spectacular effects that
are within the reach of current experiments.

\begin{acknowledgments}
We acknowledge useful discussions with S. Iblisdir, P. Massignan and
L. Tagliacozzo, and financial support from FIS2013-41757-P,
FIS2012-33642, ERC AdG OSYRIS, EU IP SIQS, EU STREP EQuaM, and
Fundaci\'o CELLEX. O.B. acknowledges support from Funda\c{c}\~{a}o
para a Ci\^{e}ncia e a Tecnologia (Portugal), namely through
programmes PTDC/POPH and projects PEst-OE/EGE/UI0491/2013,
PEst-OE/EEI/LA0008/2013, IT/QuSim and CRUP-CPU/CQVibes, partially
funded by EU FEDER, and from the EU FP7 projects LANDAUER (GA 318287)
and PAPETS (GA 323901).
\end{acknowledgments}

%
\bibliography{bib}

\end{document}